\documentclass[12pt]{article}

\usepackage[margin=1in]{geometry}  
\usepackage{graphicx}              
\usepackage{amsmath}               
\usepackage{amsfonts}              
\usepackage{amsthm}                
\usepackage{epstopdf} 
\usepackage{tikz}
\usepackage{nicefrac}
\usepackage{standalone}
\usepackage{authblk}
\usetikzlibrary{decorations.pathmorphing}
\usetikzlibrary{arrows}

\newcommand{\RR}{\mathbb{R}}      

\begin{document}

\nocite{*}

\title{Heterogeneous gain distributions in neural networks I: The stationary case}

\author[1]{Alejandro Jim\'enez Rodr\'iguez\thanks{ajimenezrodriguez1@sheffield.ac.uk}}
\author[2]{Juan Cordero Ceballos \thanks{ jccorderoc@unal.edu.co}}
\author[3]{Nestor E. Sanchez}

\affil[1]{Department of Psychology, University of Sheffield, UK}
\affil[2]{Departamento de Matem\'aticas y Estad\'istica, Universidad Nacional de Colombia-Sede Manizales, Colombia.}
\affil[3]{Departamento de F\'isica, Universidad Nacional de Colombia-Sede Manizales, Colombia.}

\maketitle

\begin{abstract}
We study heterogeneous distribution of gains in neural fields using techniques of quantum mechanics by exploiting a relationship of our model and the time-independent  Schr\"{o}dinger equation. We show that specific relationships between the connectivity kernel and the gain of the population can explain the behavior of the neural field in simulations. In particular, we show this relationships for the gating of activity between two regions (step potential), the propagation of activity throughout another region (barrier) and, most importantly, the existence of bumps in gain-contained regions (gain well). Our results constitute specific predictions that can be tested in vivo or in vitro.
\end{abstract}

\section{Introduction}

Gain modulation (GM) has been proposed as one important mechanism in the brain for neural computation and information gating \cite{mejias2014differential, salinas2000gain}. In cortical microcircuits, it is involved in attention, coordinate transformations and visual related processing in early and late stages \cite{chance2002gain, salinas2000gain, maier2010comparison}. It has also been proposed to play a major role in the propagation of epileptic seizures and the stability of the underlying networks\cite{chance2011gain, stead2010microseizures}; moreover, it has been shown that there are celular (e.g. slow afterhyperpolarization) and network (e.g. Background noise) mechanisms which can be used to control or modulate the gain of neurons in a given subpopulation \cite{higgs2006diversity, chance2002gain, mejias2014subtractive, fernandez2010gain}. Connections between GM and other mechanism of neural computation has been also revealed, for example in \cite{de2011multiplicative}, where the GM emergence as a consequence of competitive interactions between predictive coding neurons was shown.\\

GM is described as a change in the slope of the F-I relationship of a neuron, in contrast with additive changes which shift it along the current axis. These changes can be seen as changes in the sensitivity of the neuron without affecting its selectivity\cite{salinas2000gain}. The main mechanism for changing the slope of the F-I curve is the presence of background noise and a balance of excitation and inhibition \cite{chance2002gain}, something that also is known to linearize this curves\cite{ mejias2014subtractive}; however, other processes like shunting inhibition remain controversial \cite{mitchell2003shunting}. GM has been shown also to an heterogeneous property of the population, with different subpopulations (e.g. excitatory vs. inhibitory) showing different gain related behaviors \cite{higgs2006diversity, mejias2014differential} and it is natural to think, for example, in the case of attention,  that at a given time different subpupations are affected differentially . Furthermore, GM have been shown to be essential in explaining cortical shared variability in a model which includes heterogeneities in the gain and additive modulations \cite{lin2015nature}.\\
 
In this paper we propose a neural field model of heterogeneous gain modulation in a balaced network in which the excitatory and inhibitory contributions to the kernel are lumped together in and single kernel with effective excitatory effect (which have been show to be the case for E-I balance). We analize this model using methods used to study quantum wells and derive important conclusions. In particular:
\begin{itemize}
\item We show that under a specific relationship between the amount of gain modulation and the  network connectivity/E-I balance, activity can be contained or propagated betwen disjoint or neighboring neuronal regions with different gains.
\item We give a condition for the existence of bumps of activity in a gain field.
\end{itemize}
In the following sections we derive our model and its relationship with quantum wells. Then we present out main results for step potentials, barriers and wells. Finally we present our conclusions.\

\section{A neural field model for gain heterogeneities}
Consider the Amari's Neural field equation \cite{amari1977dynamics}
\begin{equation}
	\frac{\partial u}{\partial t}(x,t)=-u(x,t)+\int_{\Omega}w(x,y)f(u(y,t) - \theta)dy + I(x,t),
	\label{eq:amari}
\end{equation}
this equation models the activity of a spatially structured population of neurons as it evolves in time and space \cite{ermentrout1998neural}. $u(x,t)$ can be seen as the average membrane potential of a population at position $x$ at time $t$. $w(x,y) = w(|x - y|)$ is an isotropic coupling kernel between positions $x$ and $y$. $f:\RR\rightarrow \RR^+$, the firing rate mapping or activation function, is considered a monotonic nondecreasing function. Finally, $\theta$ is the firing threshold, above which the cell fires with a rate proportional its membrane potential. We assume $\theta = 0$.\\

\begin{figure}[!h]
\centering
\includegraphics[width=\textwidth]{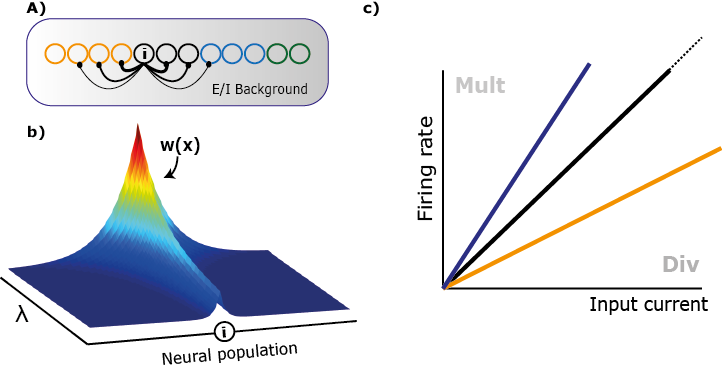}
\caption{Proposed structure of the neural network with a gain field and properties of the different neurons. a) Lateral connection for neuron $i$ with distance dependent decaying strength. The different colors represent heterogeneitiies in the network. Neurons are embeded in a noisy excitatory/inhibitory background. b) Graph representing the shape of the distribution of synaptic strengths as the parameter $\lambda$ is varied (see text) for the neuron $i$ of a). c) The heterogeneities in the network consist in different F-I modulations expresed as multiplicative(blue) or divisive (orange) changes in the slope.}
\label{fig:weights}
\end{figure}

The multiplicative modulation of the gain in the firing rate can be seen as a change in the slope of the function $f$ that depends on the position (neuron), time and even the activity of the population; therefore, asuming individual neurons with linear F-I relation, this can be expresed as
\begin{equation}
(f\circ u)(x,t) = P(x,t, u)u(x,t).
\label{eq:gainfiring}
\end{equation}
Equation (\ref{eq:gainfiring}) is called the \emph{gain field}. Let us define $P(x,t,u)$ as the difference of two important quantities: a base (mean) gain $K>0$ or excitability and spatially dependent (feedforward) modulations or fluctuations denoted by $V(x,t,u)$. Therefore,
\begin{equation}
P(x,t,u) = K - V(x,t,u),
\label{eq:pgeneral}
\end{equation}
$K$ can be thought as being a mean field or population variable, while $V$ encompasses individual neuron's diferences. In this work we assume a time and activity independent field of the form
\begin{equation}
P(x,t,u)  = P(x) = k^2 - V(x),
\label{eq:pparticular}
\end{equation}
The interaction between neurons at different positions takes the form of an exponentially decaying function and it is assumed spatially homogeneous, possitive (effectively excitatory) and isotropic
\begin{equation}
w(x,y) = w(|x-y|) =\frac{1}{2\lambda}e^{-\lambda|x-y|}, \;\; \lambda >0.\label{eq:weight}
\end{equation}
this pattern is commonly used in this kind of models to describe the empirical connection probabilities observed in some regions of the brain, like the cortex \cite{boucsein2011beyond}. Observe from figure \ref{fig:weights}(b) that as $\lambda \rightarrow +\infty$, the coupling is weaker, while, for $\lambda \rightarrow 0$, the coupling is stronger.\\
 
Under these assumptions, equation (\ref{eq:amari}) becomes
\begin{equation}
	\frac{\partial u}{\partial t}(x,t)=-u(x,t)+\frac{1}{2\lambda}\int_{\mathbb{R}}e^{-\lambda|x-y|}P(y)u(y,t)dy,
	\label{eq:amarigain}
\end{equation}
which we call the \emph{Gain Field Equation}. The structure of the population is explained in  \ref{fig:weights}(a). Our modulation of gain is multiplicaive, as oposed to additive (shifts)  \ref{fig:weights}(c).\\

It can be shown that, in the free network, $u \equiv 0$ is stable for $k^2 < \lambda^2$ and blows up for $k^2 > \lambda^2$ , however, it is not clear how the gain field affects the solution. Specifically, we are interested in three scenarios:

\begin{itemize}
\item Under which conditions can a focus of asynchronous activity be propagated from a region of high gain to a region of low gain
\item Under which conditions of the gain field can regions of high (up states) and low (down states) coexist in the same spatial domain
\item Under which conditions can this network support bumps
\end{itemize}

\section{An analytical approach for inhomogeneous gain fields}

\begin{figure}[!h]
\centering
\includegraphics[width=\textwidth]{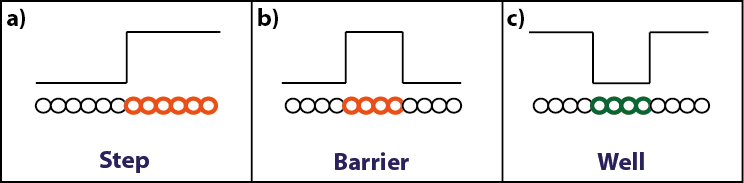}
\caption{Different gain distributions studied. a) Step gain field. Two levels of gain in the populations. b) Barrier gain field. Two homogeneous populations separated by one with lower gain. c) Synaptic well. Two lower gain populations surrounding a region with higher gain}
\label{fig:potentials}
\end{figure}

The answers to the three questions posed in the last section can be found by noticing that stationary states of the equation (\ref{eq:amarigain}) are intimately related with stationary Schr\"{o}dinger's Equation. Suppose the solutions to equation (\ref{eq:amarigain}) approaches an stationary state $u(x)$ (see appendix for a proof), then, it is given by
\begin{equation}
	u(x)=\frac{1}{2\lambda}\int_{\mathbb{R}}e^{-\lambda|x-y|}P(y)u(y)dy,
	\label{eq:amaristeady}
\end{equation}
which can be writen as
\begin{equation}
u(x)=w\ast[{f\circ{u(\cdot,t)}}](x), \label{eq:conv}
\end{equation}
where $\ast$ denotes spatial convolution and $\circ$ the function composition. Then, we can apply the Fourier Transform (FT) with respect to the spatial variable to obtain
\begin{equation}
\hat{u}=\widehat{w \ast f\circ{u}},\label{eq:ftamari}
\end{equation}
and using the fact that $\widehat{w\ast f} = \hat{w}\hat{f}$ we get
\begin{equation*}
\hat{u}(\xi)=\frac{1}{\xi^{2}+\lambda^{2}}\widehat{f\circ{u}}(\xi).
\end{equation*}
Multiplying by $\hat{w}$ on both sides and taking the inverse FT and using (\ref{eq:pparticular}) we can write
\begin{equation}
\lambda^{2}{u}(x)-u_{xx}(x)= [k^2 - V(X)]u(x),\label{e10}
\end{equation}
that is
\begin{equation} 
-\frac{d ^2 u}{dx^2}(x)+V(x)u(x)=Eu(x),\label{eq:schrodinger}
\end{equation}
with $E = k^2 - \lambda^2$. This is the time-independent Schr\"{o}dinger equation.   This relationship allow us to frame the three previous questions as inquiries about the eigenfunction $u(x)$  under three different potentials $V(x)$ (figure \ref{fig:potentials}). For each of the potentials, we are going to look for \emph{acceptable} solutions that satisfy the following conditions
\begin{enumerate}
\item $u(x)$ and $\frac{du}{dt}(x)$ should be finite (bounded and defined in all $\Omega$), because neural activity is finite by nature as well as its rate of change
\item $u(x)$ and $\frac{du}{dt}(x)$ should be continous. The continuity of $u(x)$ is necesary for the requeriment 1 and the continuity of the derivative is implied by the same form of equation (\ref{eq:schrodinger})
\item Finally, we want positive solutions given that the threshold is zero and equation (\ref{eq:amarigain}) has no mechanism for hyperpolarization. Indeed, it can be shown that, for non negative initial conditions, equation (\ref{eq:amarigain}) remains non negative (see appendix).
\end{enumerate}

\section{Results}
\subsection{The free network}
In the free network or flat potential case, we can assume $V(x) \equiv 0$ without loss of generality. This means that equation (\ref{eq:schrodinger}) becomes a simple harmonic oscillator
\begin{equation}
\frac{d ^2 u}{dx^2}(x)=-Eu(x),
\end{equation}
whose solution can be expresed as
  \begin{equation}
  u(x) = \left\{
     \begin{array}{lr}
       A\exp(iKx) + B\exp(-iKx), &  E > 0\\
       A + B, & E = 0\\
       A\exp(-Kx) + B\exp(+Kx) &,  E < 0.\\
     \end{array}
   \right.   
 \end{equation}
For some constants $A,B,C,D$ to define. The first case violates condition 3 because oscilations take positive and negative values for $A \neq 0$ and $B \neq 0$, furthermore, it implies $k^2 > \lambda^2$, which violates condition 1 for an acceptable solution because, in this case, equation (\ref{eq:amarigain}) blows up in the integral equation(see appendix for a proof). Even if they are not acceptable, they are nevetheless important in our analysis, for this reason we are going to denote those solutions as $u \equiv \hat{\infty}$. The second case satisfy all conditions trivially, so it is an acceptable solution; this implies that for $k^2 = \lambda^2$ the network presents global synchronization. The third case can only be an acceptable solution for $B = 0$, however, the stability result (see apendix) that shows the blow up, also implies the extintion of the activity for $k^2 < \lambda^2$.\\

Observe that $k^2 = \lambda^2$ corresponds to an eigenvalue of the symmetric integral operator in equation (\ref{eq:amarigain}); that explains the synchronization effect. The constant $A+B$ corresponds to the eigenfunction. It can be shown that every constant function is an eigenfunction, therefore, homogeneous initial states are unchanged when $k^2  = \lambda^2$\\
 
\begin{figure}[!htb]
\centering
\includegraphics[width=0.9\textwidth]{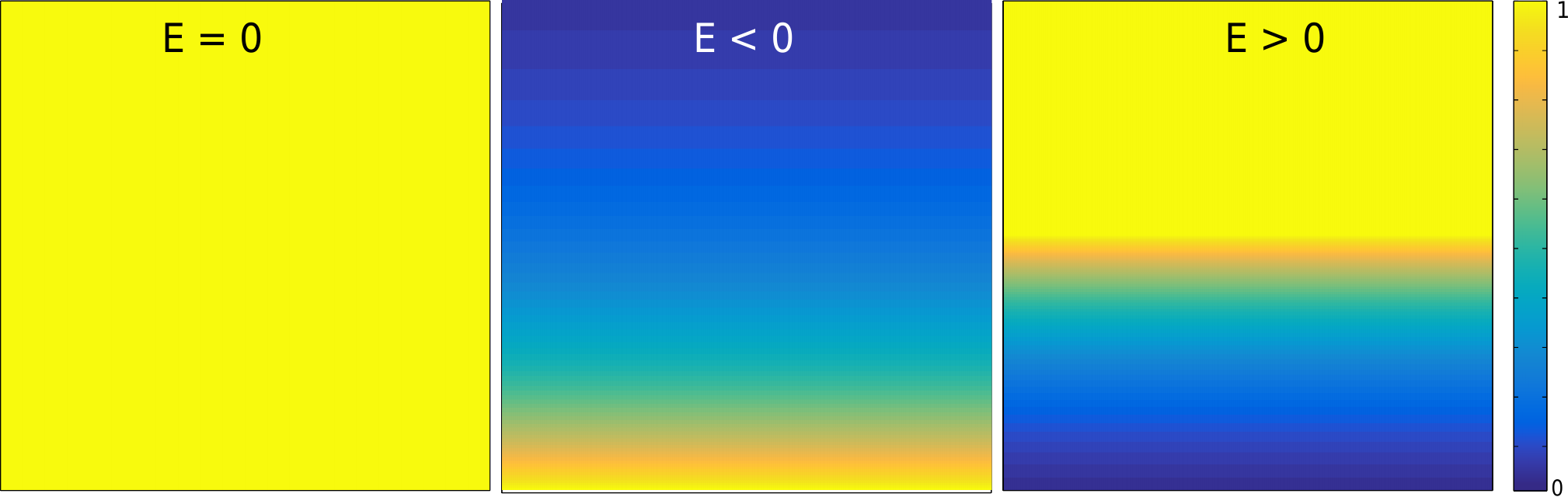}
\caption{Solutons of the gain field equation for the free network. From left to right, complete synchronization, fading out and blow up according to the theory.}
\label{fig:free}
\end{figure}

In summary, for the free case, the only acceptable solutions are a homogeneous state of synchronization $u(x) \equiv C$ when the excitability or mean gain of the population equates the spread of the lateral connectivity, and the complete extintion of the activity when the spread overpasses the mean gain. In figure (\ref{fig:free}) the two acceptable solutions are shown (b,c) along with the blow up behaviour (a).
 
\subsection{Gain step distribution}

The first gain field to study is the step potential, represented by a discontinuos function of $x$. This can be seen as the interface between two regions with different gains. Without loss of generality, we assume the potential function $V(x)$ having the following form:
\begin{align}
V(x)&=
\begin{cases}
V_0,&x \geq 0\\
0,&x<0
\end{cases}
\end{align}
That is, the $x$ axis is divided in two regions, one is kept with no modulation at all and the other has a fixed, constant modulation $V_0$. The stability results for the free network still dominate the dynamics when $E < 0$. However, for $E > 0$ we now have two more cases to analyze. The first one is when the energy is less than the potential height $E<V_0$, the other one is when the energy is greater than the potential energy $E>V_0$.  Here is how we interpret the solutions: we start as in the free case with an homogeneous initial condition $u_0(x) \equiv C$, $C > 0$ and the solutions we derive are the final distribution of neural actvities in the limit $t\rightarrow \infty$.\\

For the step potential, the equation (\ref{eq:schrodinger}) becomes
\begin{align}
-\frac{d^2 u}{dx^2}(x) &= Eu(x), &x \geq 0\\
-\frac{d^2 u}{dx^2}(x) &= (E - V_0)u(x),& x<0
\label{eq:solstep}
\end{align}

We now solve the equation in each of the regions and combine the solutions so that they satisfy the conditions stated above.

\emph{Case $1$: $E < V_0$:} In this case the eigenfunction take the form:
\begin{align}
u(x)&=
\begin{cases}
\hat{\infty},&x<0\\
De^{-\sqrt{V_0-E}x},&x>0,
\end{cases}
\end{align}
Note that, again, oscillatory solutions are not allowed and theintegral equation diverge. The constant $D$ depend on the value of the derivative at $0$.

\emph{Case $2$: $E > V_0$:} For this case, the theory predicts that the activity invades the low gain region and the eigenfunction is $u(x) = \hat{\infty}$ in the whole interval.\\

\begin{figure}[!h]
\centering
\includegraphics[width=0.8\textwidth]{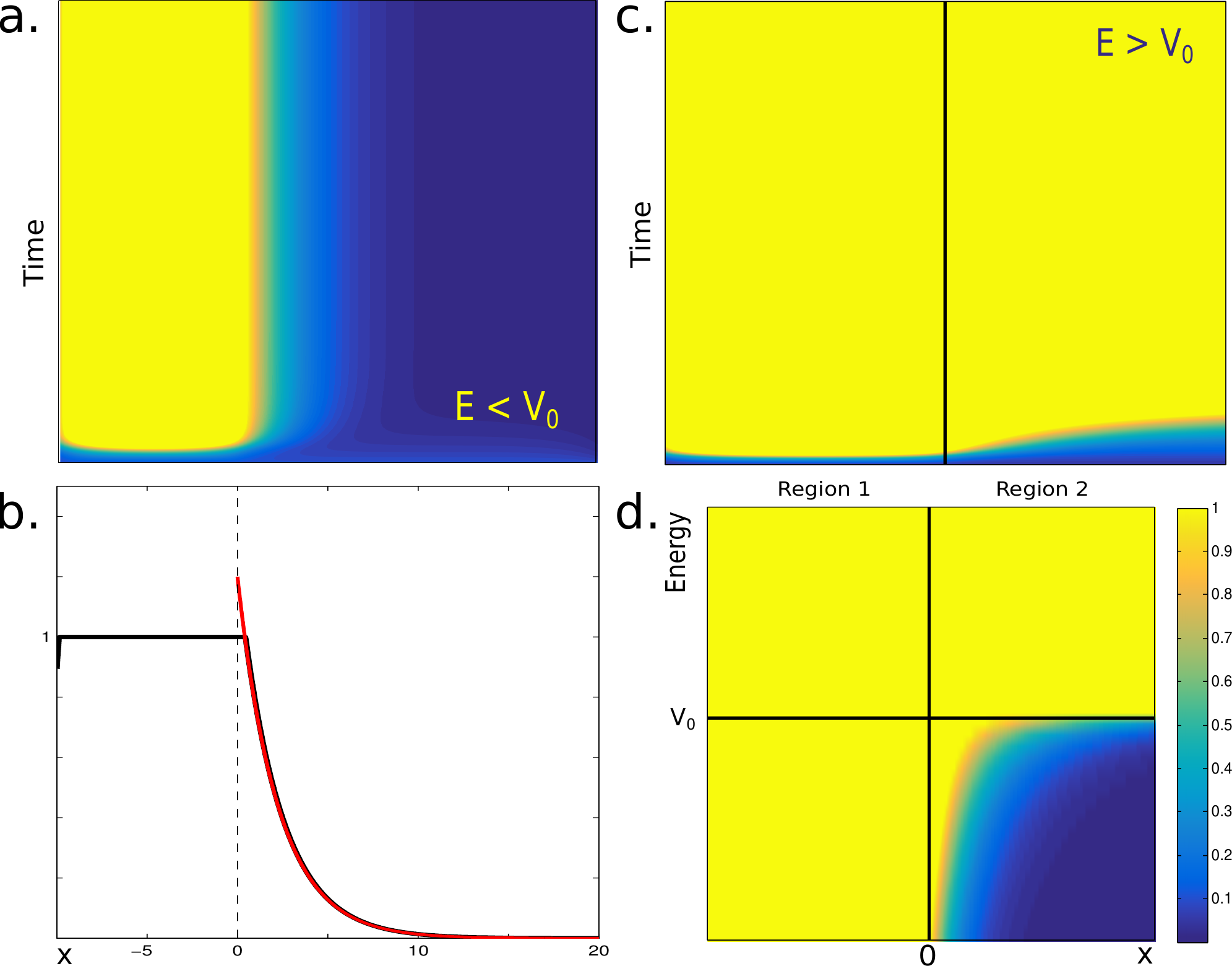}
\caption{Simulations of the neural field for the different cases for the step potential. The solution is truncated at $u(x) = 1$. a. For energy less than the height of the pulse the  activity remains in region 1. b. The profile of the solution in the stationary state and the fit of the exponential for $D = 1.2$. c. For energy higher then the height of the step, the activity invades the other regions. d. This graph shows the final profile for different values of the energy, note that the between confinement and invasion}
\label{fig:stepresult}
\end{figure}

In the figure (\ref{fig:stepresult}a),  it can be seen that, indeed, the behavior of the neural field follows precisely the predictions of the theory. For $E < V_0$, the activity is confined to the first region and there is a small probability of finding some activity in the other one. Indeed, the solution in this region, is well fitted by the exponential in (\ref{eq:solstep}) (figure (\ref{fig:stepresult}b). For $E>V_0$ there is invasion of the low gain region (figure  \ref{fig:stepresult}c). Finally, figure figure (\ref{fig:stepresult}d), shows that the final profile of the solution for different $E$ .

\subsection{Gain barriers}

\begin{figure}[!h]
\centering
\includegraphics[width=\textwidth]{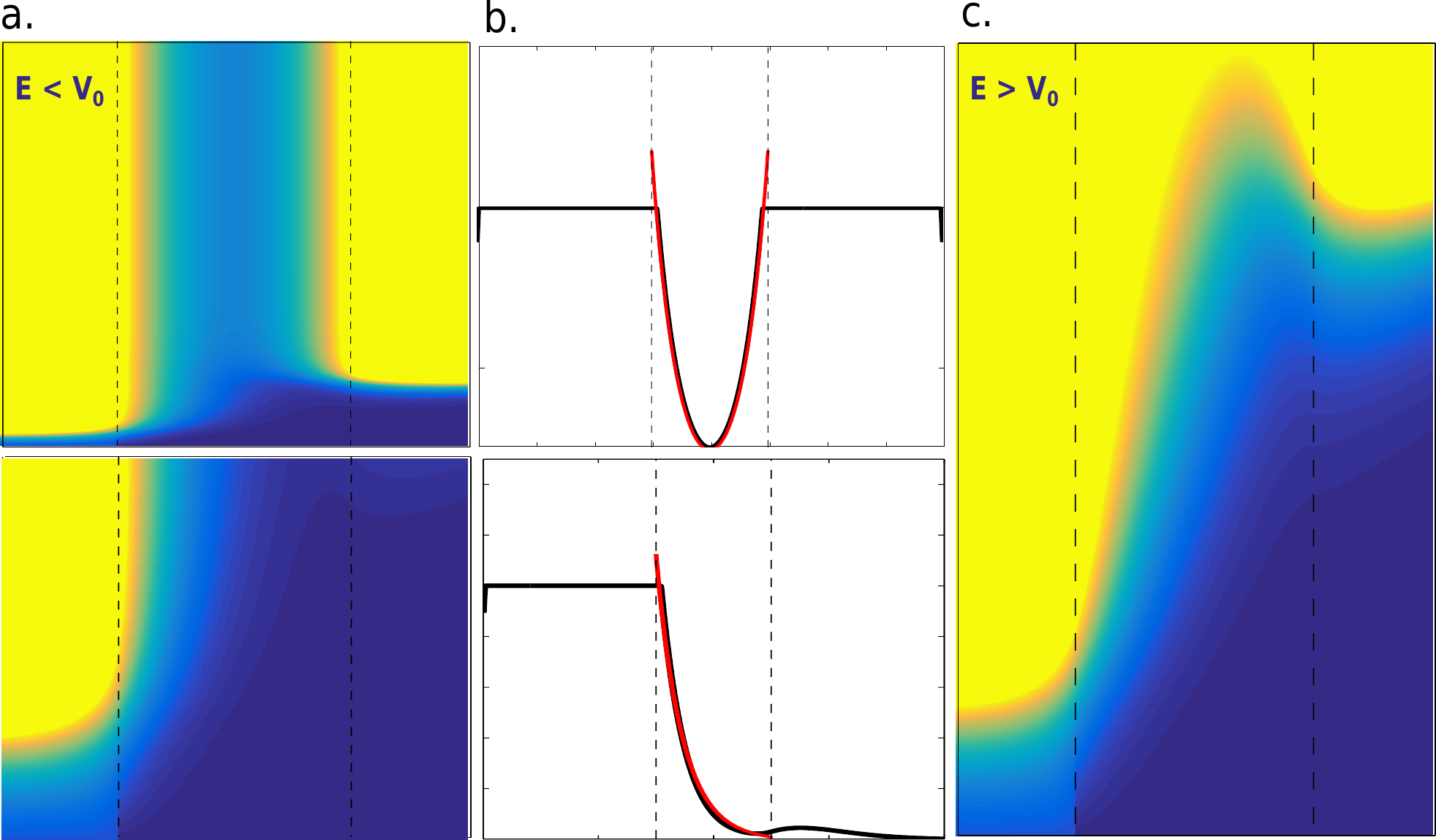}
\caption{Simulations of the neural field for the different cases for the barrier potential. Activity penetrate the barrier slowly with a profile predicted by the solution in the barrier region (1a, b). The barrier dissapears for $E> V_0$ . Red lines show fittings of the solution in the barrier region.}
\label{fig:barrier}
\end{figure}

For the gain barriers, our potential assumes a piecewise discontinuous shape with a low gain ``barrier'' separating two regions that are highly excitable, therefore, our potential takes the form
\begin{align}
V(x)=
\begin{cases}
 0,& x<-a/2, x>a/2\\
   V_0, &-a/2<x<a/2
\end{cases}
\label{equ:sqrt_pot}
\end{align}
A similar analysis as in the previous case yields the solution
\begin{align}
u(x)&=
\begin{cases}
A e^{i\sqrt{E}x} + Be^{-i\sqrt{E}x},& x < -L/2\\
F e^{-\sqrt{(V_0-E)}x} + G e^{\sqrt{(V_0-E)}x},&|x|<L/2\\
C e^{i\sqrt{E}x},& x > L/2
\end{cases}
\label{equ:sqrt_wavefunt}
\end{align}
which, gives that, outside the barrier $u = \hat{\infty}$ and penetration of the barrier for $E > V_0$ (figure  \ref{fig:barrier}).

\subsection{Gain wells}
\begin{figure}[!h]
\centering
\includegraphics[width=\textwidth]{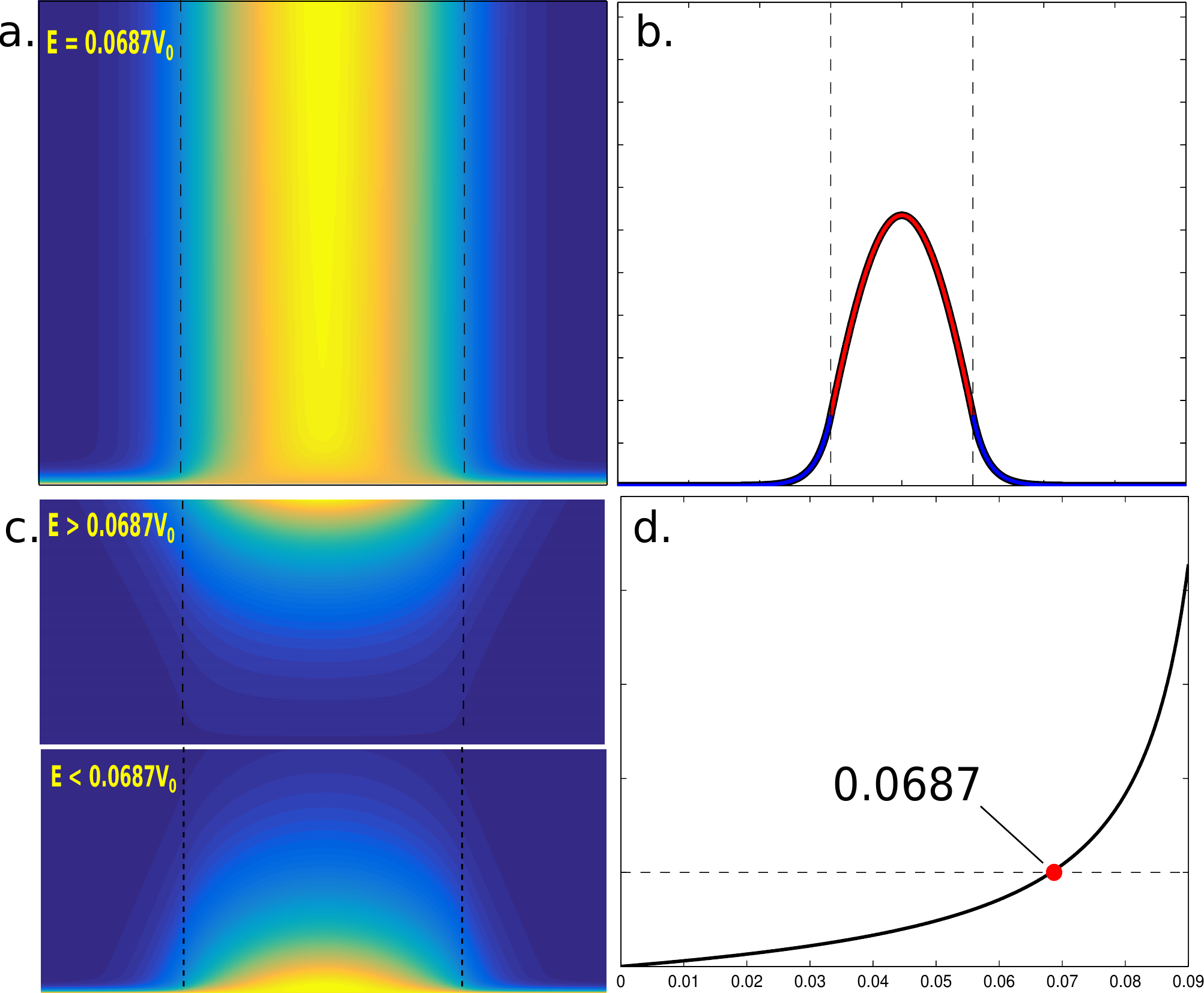}
\caption{Simulation of the neural field for the gain well. The predicted stable bump (a, b) for the eigen energy give by the first zero of (\ref{equ:perfection}), which is shown in (d). Note that for any other value of the eigenenergy, the activity either dies or blows up.}
\label{fig:well}
\end{figure}

The gain wells are modelled by square potentials, equation (\ref{equ:sqrt_pot}). 
\begin{align}
V(x)=
\begin{cases}
 V_0,& x<-a/2, x>a/2\\
  0, &-a/2<x<a/2
\end{cases}
\label{equ:sqrt_pot}
\end{align}
This distribution can be thought as a region of length $L$ of high gain surronded by two regions of less gain, and is known in quantum mechanics as a distribution of hydrogenoid type. Until now, all of our region of high gain have displayed a simple behavior; they either go to $u = \hat{\infty}$ or to $u = 0$. However, in this case, our region is a well with more subtle behavior. By following similar steps as before, we find that its solution takes the following form
\begin{align}
u(x)&=
\begin{cases}
C_{r,l} e^{\sqrt{(V_0-E)}(x)} + D_{r,l}e^{-\sqrt{(V_0-E)}(x)},&|x|\geq L/2\\
A \sin(\sqrt{E}x) + B \cos(\sqrt{E} x),&|x|<L/2
\end{cases}
\label{equ:sln}
\end{align}
Which represent  an exponential decay outside the well. Inside the well, however, we are going to follow a similar analysis as in \cite{eisberg1967fundamentals}. As before, we need to determine values for $A,B,C_{r,l},D_{r,l}$ so that this constitutes a valid solution. By enforcing the continuity of the solution and its derivatives, a system of equations is built from which it is found that the eigen-energies correspond to the zeros of the following relation.
\begin{align}
\sqrt{E \frac{L^2}{4}} \tan \left(\sqrt{E \frac{L^2}{4}}\right) =  \sqrt{(V_0 - E) \frac{L^2}{4}}
\label{equ:perfection}
\end{align}
In this case, the only period for which $u \neq \hat{\infty}$ corresponds to the first zero of this function and it is exactly the one corresponding to the bump, with shape (figure \ref{fig:well})
\begin{align}
u(x)&=
\begin{cases}
C_1 e^{-\sqrt{(V_0-E)}(|x-L/2|)},&|x|\geq L/2\\
C_2 \cos(\sqrt{E} x),&|x|<L/2
\end{cases}
\label{equ:sqrt_wavefunt}
\end{align}
 
\section{Conclusion}
We developed a new approach to the analysis in gain modulation in structured neural networks that includes heterogeneities, the power of which has been illustrated by making specific predictions that can be tested about the propagation and containment of activity, and, particularly, the existence of bumps in an effectively excitatory network. It has been shown previously that spiking neural networks can support bumps \cite{laing2001stationary} and our predictions are testable in this kind of more biologically plausible networks by modulating the slope if the F-I curve of specific subpopulation as shown in \cite{chance2002gain}, in vivo, and in \cite{chance2011gain, mejias2014subtractive} computationally. This is subject of ongoing work by the authors. Furthermore, the analysis performed can be extended to different shapes of the potential and non-stationary distributions and it can benefit from the important advances made already in quantum dynamics in the study of quantum wells.\\ 

The posibility of supporting bumps without any additional plasticity mechanism is an interesting finding that can be related with the stability of sensory representations and induced short therm memory in downstream regions without explicit need of other means of adaptation. In general, the energy, defined as $E = k^2 - \lambda^2$  establishes a relationship between E-I balance, spread of connectivity and the slope of the F-I curve. In the case of step potentials, in which we are interested in invasion of activity in one region to activity in other region, the population gain ($k$) can be externally manipulated to put the energy in the appropiate regime to gate propagation, without changing the effective synaptic input. In the case of the barrier potential, further study  is necessary to determine the rate of penetration (flow) across the barrier; this result could be of interest in studing epileptic microseizures which have focal origins that can be propagated to regions of different properties \cite{stead2010microseizures}.

\section{Acknowledgements}
\label{acknowledgements}
This project, with title  ``An\'alisis te\'orico - Experimental de un modelo de campo Neural usando t\'ecnicas de la mec\'anica cu\'antica", was funded by the ``Convocatoria del programa nacional de proyectos para el fortalecimiento de la investigaci\'on, la creaci\'on y la innovaci\'on de posgrados de la universidad Nacional de Colombia 2013 - 2015", code number 19375, of the Universidad Nacional de Colombia, sede Manizales. The authors want to thank professor Carlos Vargas for the discussions and suggestions held about the physics of quantum wells.

\bibliographystyle{plain}

\bibliography{paper}

\end{document}